\documentclass[12pt]{article}
\usepackage{tikz}
\usetikzlibrary{matrix,arrows,decorations.pathmorphing,decorations.pathreplacing,decorations.markings,backgrounds,positioning,fit,petri}
\usepackage{jheppub}
\usepackage{amsmath,amssymb,euscript,array,mathrsfs}

\usepackage{epsfig}

\def\XXint#1#2#3{{\setbox0=\hbox{$#1{#2#3}{\int}$}
     \vcenter{\hbox{$#2#3$}}\kern-.5\wd0}}

\newcommand{\Tr}{\operatorname{Tr}}

\def\BH{{\cal H}}

\def\Tr{{\rm Tr}}

\def\Dbarslash{\,\,{\raise.15ex\hbox{/}\mkern-12mu {\bar D}}}
\def\Dslash{\,\,{\raise.15ex\hbox{/}\mkern-12mu D}}
\def\delslash{\,\,{\raise.15ex\hbox{/}\mkern-9mu \partial}}
\def\delbarslash{\,\,{\raise.15ex\hbox{/}\mkern-9mu {\bar\partial}}}

\def\bra#1{\langle #1|}
\def\ket#1{| #1\rangle}

\newcommand{\EQ}[1]{\begin{equation*}\begin{split} #1 \end{split}\end{equation*}}

\def\LCATH{\raisebox{-2pt}{\begin{tikzpicture}[scale=0.13]
\draw (0,0) ellipse (1.8cm and 1.2cm);
\draw [-] (1,0) -- (2.3,0);
\draw [-] (-1,0) -- (-2.3,0);
\draw [-] (1,-0.2) -- (2.2,-0.5);
\draw [-] (-1,-0.2) -- (-2.2,-0.5);
\draw [-] (1,0.2) -- (2.2,0.5);
\draw [-] (-1,0.2) -- (-2.2,0.5);
\filldraw[black] (0.4,0.3) circle (0.1cm);
\filldraw[black] (-0.4,0.3) circle (0.1cm);
\draw[-] (-0.5,-0.6) to[out=-30,in=210] (0.5,-0.6);
\draw[-] (0.8,1.1) -- (1.1,1.8) -- (1.4,0.8);
\draw[-] (-0.8,1.1) -- (-1.1,1.8) -- (-1.4,0.8);
\end{tikzpicture}
}\!\!}

\def\DCATH{\raisebox{-2pt}{\begin{tikzpicture}[scale=0.13]
\draw (0,0) ellipse (1.8cm and 1.2cm);
\draw [-] (1,0) -- (2.3,-0.5);
\draw [-] (-1,0) -- (-2.3,-0.5);
\draw [-] (1,-0.2) -- (2.2,-1);
\draw [-] (-1,-0.2) -- (-2.2,-1);
\draw [-] (1,0.2) -- (2.2,0);
\draw [-] (-1,0.2) -- (-2.2,0);
\draw[-] (0.2,0.1) -- (0.6,0.5);
\draw[-] (0.2,0.5) -- (0.6,0.1);
\draw[-] (-0.2,0.1) -- (-0.6,0.5);
\draw[-] (-0.2,0.5) -- (-0.6,0.1);
\draw[-]  (-0.5,-0.6) to[out=30,in=150] (0.5,-0.6);
\draw[-] (0.8,1.1) -- (1.1,0.6) -- (1.4,0.8);
\draw[-] (-0.8,1.1) -- (-1.1,0.6) -- (-1.4,0.8);
\end{tikzpicture}
}\!\!}

\title{Schr\"odinger's Cat and the Firewall}
\author{Timothy J. Hollowood}
\affiliation{Department of Physics, Swansea University,
Swansea, SA2 8PP, UK.}
\emailAdd{t.hollowood@swansea.ac.uk}

\abstract{It has been argued that when black holes are treated as quantum systems there are implications at the horizon and not just the singularity. Infalling observers will meet a firewall of high energy quanta. We argue that the question of whether an observer falling into a black hole experiences a smooth horizon or a firewall is identical to the question of whether Schr\"odinger's cat is either in a definite state, alive or dead, or in a superposition of the two. Since experience with real macro-systems indicate the former, the black hole state vector is seen to describe a set of decoherent alternatives each with a smooth horizon and the entanglement puzzle is thereby side stepped.
\vspace{1cm}
\begin{center}
{\bf Essay written for the Gravity Research Foundation 2014 Awards for Essays on Gravitation}\\

\vspace{1cm}
{\bf ---submitted 24th March 2014---}  \end{center}

}

\notoc
\begin{document}

\maketitle

\newpage

\noindent Black holes are are arguably the most fascinating objects in the universe because the interplay between gravity and quantum mechanics cannot be avoided. If we understand how the two great pillars of 20 century physics find accommodation here, then surely we will have understood something profound about how the universe works.

While it clear that general relativity breaks down at the singularity and quantum effects are unavoidable what is more interesting is that gravity and quantum mechanics seem to have some important effects near the horizon where the curvature is not large and we would expect to able to formulate quantum field theory in a fixed spacetime background \cite{Almheiri:2012rt}. The question is what will happen to Schr\"odinger's cat should she fall into a very large black hole?

First of all let us focus on the region near the horizon and in particular on the states of the Hawking radiation $\ket{n}$ with energy $E_n$ that have just been emitted by the black hole of mass $M$ in the state $\ket{i}$. After unitary evolution, the radiation will be entangled with  the black hole \cite{Verlinde:2012cy} 
\EQ{
&\ket{i}\ket{0}\longrightarrow \sum_nc_n\ket{\psi^i_n}\ket{n}\ ,\\
\text{with}&\qquad\bra{n}m\rangle=\bra{\psi^i_n}\psi^i_m\rangle=\delta_{nm}\ ,
\label{p22}
}
where $\ket{\psi_n}$ are states of the black hole of mass $M-E_n$. Because the Hilbert space of the radiation---with some UV cut off---is much smaller than that of the black hole, standard statistical reasoning implies that coefficients take the form
$|c_n|^2\approx e^{-\beta E_n}/Z$, where $\beta$ is an effective inverse temperature given by $dS/dM=8\pi M$, with $S$ the black hole entropy. So the reduced density matrix of the external radiation is just a thermal ensemble of Hawking radiation: 
\EQ{
\rho_\text{rad}\approx\frac1Z\sum_ne^{-\beta E_n}\ket{n}\bra{n}\ ,\qquad Z=\sum_ne^{-\beta E_n}\ .
}

For each $i$, the states of the black hole $\ket{\psi^i_n}$, are naturally interpreted as states of the radiation propagating on the smooth geometry continued across the horizon and the quantum state above is just the Unruh vacuum. 

In some respects one could say that internal geometry emerges out of the quantum states of the black hole that are entangled with the external radiation. This is literally what happens in the context of the AdS/CFT correspondence where the whole geometry is emergent \cite{Papadodimas:2013jku}. 

However, this nice picture is spoiled as the black hole emits more and more radiation \cite{Verlinde:2012cy}. Taking account of the radiation that has previously been emitted and is now causally separated from the black hole, the overall state vector becomes
\EQ{
&\sum_is_i\ket{i}\ket{0}\ket{\Phi_i}\longrightarrow \sum_is_i\Big(\sum_nc_n\ket{\psi^i_n}\ket{n}\Big)\ket{\Phi_i}\ ,\\
&~~~~~~~~~\text{with}\qquad\bra{i}j\rangle=\bra{\Phi_i}\Phi_j\rangle=\delta_{ij}\ .
\label{p23}
}
Note that only the black hole and near radiation are coupled by the Hamiltonian so that the states of the early radiation $\ket{\Phi_i}$ are just inert spectators. 

But now the fragile entanglement between the states of the black hole and near radiation has been damaged by entanglement with the early radiation as dictated by the purity of the overall state: entanglement is famously monogamous and will not be shared. The implication is that if one associates the state to a single background geometry then state of the radiation across the horizon is no longer the Unruh vacuum and is likely to be a state with quanta excited near the horizon in the form of a deadly firewall of radiation \cite{Almheiri:2012rt}. 

It seems that Schr\"odinger's cat will be microwaved as she crosses the horizon.

\vspace{0.2cm}
\begin{center}
{\tiny***}
\end{center}
\vspace{0.2cm}

\noindent The analysis so far just assumes unitary evolution of the state vector; however, we know that the ad hoc rules of the Copenhagen interpretation are needed in order to  describe successfully the phenomenology of macro-systems.

A simple qubit version of the Schr\"odinger's cat experiment has time evolution
\EQ{
\big(c_+\ket{z^+}+c_-\ket{z^-}\big)\ket{\LCATH}\longrightarrow c_+\ket{z^+}\ket{\LCATH}+c_-\ket{z^-}\ket{\DCATH}\ ,
}
where $\ket{\DCATH}$ and $\ket{\LCATH}$ are macro-states corresponding to a dead/alive cat. The ad hoc rules suggest that we should collapse the final state vector to either $\ket{z^+}\ket{\LCATH}$ or $\ket{z^-}\ket{\DCATH}$ based on the fact that these are macroscopically distinct. 


Ad hoc rules are by their nature not helpful in unfamiliar situations a like black hole.
One way to do better is to literally interpret the reduced density matrix of the cat  as a statistical mixture of actual, or ontic, states 
\EQ{
&\rho_\text{cat}=|c_+|^2\ket{\LCATH}\bra{\LCATH}+|c_-|^2\ket{\DCATH}\bra{\DCATH}\ ,\\
&~\text{ontic states}=\begin{cases} \ket{\LCATH} & \text{prob}=|c_+|^2\\ \ket{\DCATH} & \text{prob}=|c_-|^2\ ,\end{cases}
}
i.e.~the cat is in a definite state either alive or dead with probabilities that yield the Born rule. This is the essence of {\it modal\/} quantum mechanics, states have different {\it modalities\/}: the overall state vector and then the ontic state associated to the perspective of a sub-system in the form of one of the eigenvectors of its reduced density matrix.

Although containing the essence, the presentation above is rather too simple: one should include some element of coarse graining and an environment, so that the states of the cat are no longer pure, as well as address the issues of time dependence and degeneracy \cite{Hollowood:2013cbr,long}.  Local observers only have access to local degrees-of-freedom and have finite resolution, so more generally
the relevant density matrix is associated to a set of coarse-grained local observables ${\cal A}$ defined so that \cite{Jaynes2}
(i) $\text{Tr}(\rho{\cal O}_I)=\bra{\Psi}{\cal O}_I\ket{\Psi}$ for all ${\cal O}_I\in{\cal A}$; and
(ii) $\rho$ has maximal von~Neumann entropy $S=-\text{Tr}(\rho\log\rho)$. 
It takes the form
\EQ{
\rho=\frac1Z\exp\Big[-\sum_I\mu_I{\cal O}_I\Big]\ ,
}
which identifies the ontic states as the eigenvectors of the modular Hamiltonian $\sum_I\mu_I{\cal O}_I$. If the set ${\cal A}$ did consist of a complete set of observables on a tensor product factor $\BH_A\otimes\BH_E$, then $\rho=\rho_A\otimes\text{Id}/d_E$, and so we recover a description in terms of the reduced density matrix $\rho_A$.

\vspace{0.6cm}
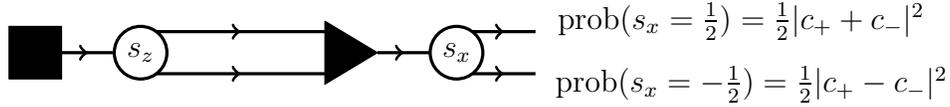
\begin{figure}[h]
\begin{center}
\begin{tikzpicture}[scale=0.7,decoration={markings,mark=at position 0.5 with {\arrow{>}}}]
\filldraw[black] (-0.5,-0.5) -- (0.5,-0.5) -- (0.5,0.5) --(-0.5,0.5) -- (-0.5,-0.5);
\draw[very thick] (2,0) circle (0.5cm);
\node at (2,0) (a1) {$s_z$};
\draw[very thick,postaction={decorate}] (0.5,0) -- (1.5,0);
\begin{scope}[xshift=6cm,yshift=0cm]
\filldraw[black] (-0.5,0.6) -- (0.5,0) -- (-0.5,-0.6) -- (-0.5,0.6);
\end{scope}
\draw[very thick,postaction={decorate}] (2.3,0.4) -- (5.5,0.4);
\draw[very thick,postaction={decorate}] (2.3,-0.4) -- (5.5,-0.4);
\draw[very thick] (8,0) circle (0.5cm);
\node at (8,0) (a1) {$s_x$};
\draw[very thick,postaction={decorate}] (6.5,0) -- (7.5,0);
\draw[very thick,postaction={decorate}] (8.3,0.4) -- (9.5,0.4);
\draw[very thick,postaction={decorate}] (8.3,-0.4) -- (9.5,-0.4);
\node at (13.4,0.6) (b2) {$\text{prob}(s_x=\frac12)=\frac12|c_++c_-|^2$};
\node at (13.6,-0.6) (b3) {$\text{prob}(s_x=-\frac12)=\frac12|c_+-c_-|^2$};
\end{tikzpicture}
\end{center}
\caption{\small A experiment involving a qubit produced by a source in the state $\ket{\psi}=c_+\ket{z^+}+c_-\ket{z^-}$. The qubit state is split by a polariser into eigenstates of $s_z$ which are then combined to give the original state vector $\ket{\psi}$. A measurement of $s_x$ is then made yielding the probabilities as shown.}
\label{f1}
\end{figure}
A more sophisticated analysis of realistic systems shows that the ontic states are precisely those that the Copenhagen interpretation would collapse to, but in the modal approach the collapse is just a piece of---fundamentally unnecessary---spring cleaning that removes unrealised possibilities. 
The modal approach is therefore able to cleanly identify the distinct measurement outcomes and quantify their decoherence \cite{long}. 

To illustrate this further, take the simple quantum system illustrated in fig.~\ref{f1}. A source produces a qubit in the state 
\EQ{
\ket{\psi}=c_+\ket{z^+}+c_-\ket{z^-}\ ,
}
which is passed to a polariser that selects eigenstates of $s_z$. The two possible selections are then combined, so this part is redundant for now, before finally another polariser measures $s_x$. 

\vspace{0.6cm}
\begin{figure}[h]
\begin{center}
\begin{tikzpicture}[scale=0.7,decoration={markings,mark=at position 0.5 with {\arrow{>}}}]
\filldraw[black] (-0.5,-0.5) -- (0.5,-0.5) -- (0.5,0.5) --(-0.5,0.5) -- (-0.5,-0.5);
\draw[very thick] (2,0) circle (0.5cm);
\node at (2,0) (a1) {$s_z$};
\draw[very thick,postaction={decorate}] (0.5,0) -- (1.5,0);
\begin{scope}[xshift=6cm,yshift=0cm]
\filldraw[black] (-0.5,0.6) -- (0.5,0) -- (-0.5,-0.6) -- (-0.5,0.6);
\end{scope}
\draw[very thick,postaction={decorate}] (2.3,0.4) -- (5.5,0.4);
\draw[very thick,postaction={decorate}] (2.3,-0.4) -- (5.5,-0.4);
\draw[very thick] (8,0) circle (0.5cm);
\node at (8,0) (a1) {$s_x$};
\draw[very thick,postaction={decorate}] (6.5,0) -- (7.5,0);
\draw[very thick,postaction={decorate}] (8.3,0.4) -- (9.5,0.4);
\draw[very thick,postaction={decorate}] (8.3,-0.4) -- (9.5,-0.4);
\filldraw[black] (3.9,0.4) circle (0.2cm);
\filldraw[black] (3.9,-0.4) circle (0.2cm);
\draw[very thick,densely dashed] (3.9,0.4) -- (3.5,0) -- (3.5,-1.5);
\draw[very thick,densely dashed] (3.9,-0.4) -- (3.7,-0.8) -- (3.7,-1.5);
\begin{scope}[xshift=3.6cm,yshift=-2cm]
\filldraw[black] (-0.5,-0.5) -- (0.5,-0.5) -- (0.5,0.5) --(-0.5,0.5) -- (-0.5,-0.5);
\end{scope}
\draw[very thick,postaction={decorate}] (3.6,-2.5) -- (3.6,-3.5);
\node at (10.8,-2.7) (b4) {$\rho_1$ (i.e.~$\ket{z^+}$ or $\ket{z^-}$ with $\text{prob}=|c_\pm|^2$)};
\draw[->] (6.2,-2.3) -- (7,-0.2);
\node at (3.6,-3.9) (b3) {qubit 2};
\node at (12.4,0.6) (b2) {$\text{prob}(s_x=\frac12)=\frac12$};
\node at (12.6,-0.6) (b3) {$\text{prob}(s_x=-\frac12)=\frac12$};
\end{tikzpicture}
\end{center}
\caption{\small A variation of the experiment in fig.~\ref{f1} including ``which way" detectors that are used to prepare the state of a second qubit so that the total state of both qubits is the entangled state $\ket{\Psi}$. A measurement of $s_x$ is then made on qubit 1. From the point of view of qubit 1, its state before measurement is the reduced density matrix $\rho_1$ effectively, therefore, a decoherent mixture of either $\ket{z^\pm}$ with probabilities $|c_\pm|^2$. The measurement of $s_x$ for either of $\ket{z^\pm}$ yields $\pm\frac12$ with probabilities $\frac12$.}
\label{f2}
\end{figure}
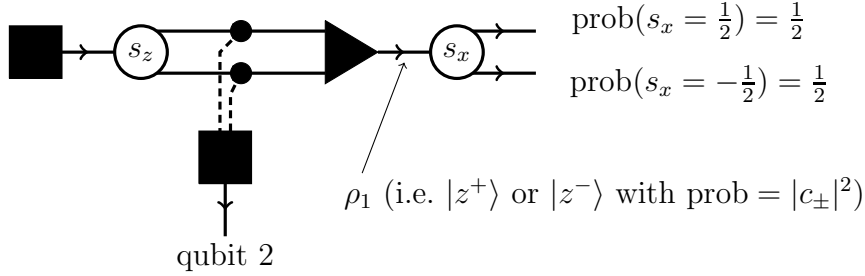

Now consider the variant in fig.~\ref{f2} which collects ``which way" information as indicated and prepares a second qubit so that the total state is now entangled:
\EQ{
\ket{\Psi}=c_+\ket{z^+z^+}+c_-\ket{z^-z^-}\ .
}
The local state of qubit 1 is now the reduced density matrix
\EQ{
\rho_1=\Tr_2\ket{\Psi}\bra{\Psi}=|c_+|^2\ket{z^+}\bra{z^+}+|c_-|^2\ket{z^-}\bra{z^-}\ ,
}
rather than the state $\ket{\psi}$,
which we can interpret in the modal way as saying that the ontic state of qubit 1 is either $\ket{z^\pm}$ with probabilities $|c_\pm|^2$. For both ontic states, the resulting measurement of $s_x$ then yields $\pm\frac12$ with probabilities $\frac12$. It is important that any local manipulation made on causally separated qubit 2 cannot change $\rho_1$ and therefore affect the ontic state of qubit 1 even if the ``which way" information is erased or some component of the spin measured.

From the perspective of qubit 1, causally separated from qubit 2 as it is, in the language of the Copenhagen interpretation, its state vector can effectively be collapsed to one of the pair $\ket{z^\pm}$ with probabilities $|c_\pm|^2$. 
However, to some non-local observation that has access to both qubits, the appropriate state would still be $\ket{\Psi}$ emphasising that in modal approaches the relevant state depends on the observer's perspective.

In this modal interpretation locality is paramount and there is no real collapse. Quantum states are local properties but for a selection of macro-systems embedded in an overall environment these local descriptions knit together to give a classical whole \cite{long}.

\vspace{0.2cm}
\begin{center}
{\tiny***}
\end{center}
\vspace{0.2cm}

\noindent Now we return to the firewall paradox to make a very simple point. An infalling observer crossing the horizon 
is causally separated from the early Hawking radiation and so from its perspective the local state is the reduced density matrix
\EQ{
\rho_\text{infall}=\sum_i|s_i|^2\ket{\varphi^i}\bra{\varphi^i}\qquad\text{where}\qquad\ket{\varphi^i}=\sum_nc_n\ket{\psi_n^i}\ket{n}\ .
}
The eigenvectors $\ket{\varphi^i}$, the ontic states, are effectively the collapsed states of the Copenhagen interpretation relevant for an infalling observer. We have already argued that $\ket{\varphi^i}$, for each $i$, can be viewed as the Unruh vacuum on a smooth geometry across the horizon. Since the two sub-systems---early radiation and the rest---are causally disconnected, the $\ket{\varphi^i}$ are completely decoherent and can never interfere like the states of qubit 1 in our simple model where qubit 2 plays the role of the early radiation.

The implication is that from the perspective of the infalling observer, the total state vector does not describe radiation states on a single smooth geometry but rather describes an ensemble of possible states, each of which is associated to a smooth geometry. Notice that the entanglement problem is solved because the entanglement required to have a smooth geometry and that required to have purity of the overall state refer to different states; namely $\ket{\varphi^i}$, for some $i$, and $\sum_is_i\ket{\varphi^i}\ket{\Phi_i}$, respectively. 

So there's good news and bad news for Schr\"odinger's cat: if we throw her into a very large black hole she will escape being microwaved at the horizon and endure but only until her ultimate demise at the singularity.

\vspace{0.2cm}
\begin{center}
{\tiny******}
\end{center}
\vspace{0.2cm}

\noindent I am supported in part by the STFC grant ST/G000506/1.

\end{document}